\documentclass[11pt]{article}
\usepackage[T1]{fontenc}
\usepackage[utf8]{inputenc}
\usepackage{lmodern}
\usepackage[margin=1in]{geometry}
\usepackage{amsmath,amssymb,amsthm,mathtools,bm}
\usepackage{microtype,booktabs,tabularx,array}
\usepackage{enumitem,needspace}
\usepackage[round,authoryear]{natbib}
\usepackage[hidelinks]{hyperref}
\usepackage{fancyhdr}
\hypersetup{pdftitle={Truthful and Admissible Principal--Agent Testing with Nuisance Parameters},pdfauthor={Research draft},pdfsubject={E-values, elicitation, nuisance parameters, admissibility, composite hypothesis testing}}
\numberwithin{equation}{section}
\newtheorem{theorem}{Theorem}[section]
\newtheorem{proposition}[theorem]{Proposition}
\newtheorem{lemma}[theorem]{Lemma}
\newtheorem{corollary}[theorem]{Corollary}
\theoremstyle{definition}
\newtheorem{definition}[theorem]{Definition}

\theoremstyle{remark}
\newtheorem{remark}[theorem]{Remark}
\newcommand{\R}{\mathbb R}
\newcommand{\E}{\mathbb E}
\newcommand{\PP}{\mathcal P}
\newcommand{\QQ}{\mathcal Q}
\newcommand{\X}{\mathcal X}
\newcommand{\ind}{\mathbf 1}
\newcommand{\one}{\mathbf 1}
\newcommand{\sgn}{\operatorname{sgn}}
\newcommand{\KL}{\operatorname{KL}}
\newcommand{\Ber}{\operatorname{Ber}}

\newcommand{\argmax}{\operatorname*{arg\,max}}
\newcommand{\ip}[2]{\left\langle #1,#2\right\rangle}
\newcommand{\norm}[1]{\left\lVert #1\right\rVert}
\newcommand{\eps}{\varepsilon}

\title{\textbf{Truthful and Admissible Principal--Agent Testing\\with Nuisance Parameters}}
\author{Aaditya Ramdas \\ \texttt{aramdas@stanford.edu}}
\date{\today}

\begin{document}
\maketitle


\begin{abstract}
Statistical contracts reward experimenters according to the evidence produced by a costly study, so the reward rule influences both who participates and what they report. Principal--agent testing theory explains how to prevent ineffective proposals from being profitable. We ask when such contracts can also elicit honest parameter reports if the decision concerns a mean effect but the variance or distributional shape is unknown. Under risk neutrality and a fixed participation cost, we construct menus of e-values --- nonnegative normalized rewards with expected value at most one under every null --- such that (i) agents who know they are null have no incentive to participate, (ii) every alternative type earns positive expected profit and is thus incentivized to participate, (iii) truthful reporting uniquely maximizes expected their reward,  (iv) and each contract is admissible: its payouts cannot be pointwise improved while retaining null validity. For Gaussian observations, we treat separately agents who know both mean and variance and agents who know only their mean; in the latter case, truthfulness holds at every variance without requiring a nuisance prior. Multi-sample extensions preserve these incentives and can reward accumulating evidence. Beyond Gaussian models, we show that two independent nonnegative observations suffice to elicit any finite mean above a positive null threshold, without any further distributional knowledge, but one observation cannot combine strict mean elicitation with individual admissibility on this class. Finally, we distinguish the existence of useful tests from the identification needed for truthful contracts. Boundedness is not required, and finitely many null constraints do not by themselves guarantee admissible elicitation.
\end{abstract}

\section{Introduction}
\label{sec:intro}

Consider a buyer evaluating a technology that promises to improve average performance. The developer has private information about the technology, but the buyer requires experimental evidence before offering a reward. The developer must decide whether an experiment is worth its cost and what effect size to claim beforehand. The buyer would like useful technologies to enter, ineffective technologies not to profit in expectation, and claims to reflect what the developer actually knows. A statistical rule therefore has two roles: it evaluates the data produced by an experiment, and it shapes the incentives to undertake that experiment and disclose private information.

This is a problem in \emph{statistical contract theory}: the design of agreements whose rewards depend on statistical evidence and whose participants act strategically. The decision-maker offering the agreement and reward is the \emph{principal}; the prospective experimenter with private information is the \emph{agent}. Unlike an ordinary testing problem with a given sample, the contract must account for the agent's decision to participate. Unlike ordinary belief elicitation, it must also protect the principal against rewarding claims that fail a statistical requirement. Work on strategic testing makes these two considerations part of the same design problem \citep{tetenov2016,bates2022}.

\subsection{From statistical tests to contracts}

\citet{tetenov2016} shows how to discourage undesirable proposals by tying the testing threshold to the cost of an experiment and the benefit of approval. \citet{bates2022} develop a more flexible principal--agent framework in which the agent selects from a menu of data-dependent profit licenses. Their central characterization connects contracts that make null proposals unprofitable to \emph{e-values}.
Their reward is not a fixed prize for crossing a significance threshold: different experimental outcomes can earn different rewards.

The connection has a simple interpretation. Let $C>0$ be the cost of a specified experiment, and let $L_r(X)$ be the nonnegative reward after the agent selects report $r$ and the experiment produces data $X$. The report is chosen before sampling, and the cost does not depend on the report. Write $\PP$ for the collection of distributions under which the proposal fails the principal's requirement. Preventing positive expected profit under every such distribution amounts to requiring
\begin{equation*}
E_r(X):=\frac{L_r(X)}{C}\geq0,
\qquad
\sup_{P\in\PP}\E_P E_r(X)\leq1
\quad\text{for every offered report }r.
\end{equation*}
This is exactly the defining condition of an e-value. Large values provide evidence against the null: by Markov's inequality, the probability of $E_r\geq1/\alpha$ is at most $\alpha$ under any null distribution. The corresponding monetary reward cannot exceed the participation cost in expectation under a null distribution, although a realization providing strong evidence can earn a large reward. A risk-neutral agent's expected net payoff is $C(\E_P E_r-1)$. Thus the same expectation governs statistical validity and the incentive to enter.

Subsequent work uses participation to learn about agents' private beliefs \citep{bates2023} and extends error guarantees to risk-sensitive agents \citep{shi2024}. Closest to our reporting objective, \citet{shi2025} construct separating menus that elicit a private prior probability over two possible states and match the testing performance of a decision-maker who knows that probability. Their analysis explicitly connects contract design with proper scoring rules---reward rules that make truthful reporting optimal. The distinction between deterring null entry and eliciting private information is therefore already present in this literature; our question is how to accomplish both when the private information concerns a sampling-model parameter and nuisance features remain unrestricted and possibly infinite dimensional.

\subsection{The nuisance-parameter question}

Our starting example is
\begin{equation}
X\sim N(\mu,\sigma^2),\qquad
H_0:\mu\leq 0,\qquad H_1:\mu>0,\qquad \sigma>0.
\label{eq:gaussian-model}
\end{equation}
The principal cares whether the average effect $\mu$ is positive, not whether the observations are more or less variable. Nevertheless, the variance changes the distribution of evidence and hence the expected reward of each contract. It is a \emph{nuisance parameter}: irrelevant to which hypothesis is true, but relevant to how the data behave. The general framework of \citet{bates2022} already accommodates composite nulls such as this one. Our contribution is to construct contracts with additional reporting guarantees.

We distinguish two information regimes. In the first, the agent knows both $\mu$ and $\sigma$; the principal would like to solicit both (from a non-null agent) without creating an incentive to exaggerate the effect or misrepresent the noise. In the second regime, the agent knows $\mu$ but not $\sigma$. Asking for a full distribution would then require information the agent does not possess. Instead, we require the true mean to be the uniquely optimal report for a non-null agent \emph{regardless of the variance}. A participating non-null agent is thus incentivized to report their mean honestly, without knowing which variance applies or adopting a particular prior over variances.

In both regimes, we naturally maintain that null agents must have no profitable report, but we also insist that every positive-mean agent must strictly benefit from participating. We finally require each contract to be \emph{admissible}: no other valid e-value can increase its payout pointwise, with a strict increase on a nonnegligible set. This rules out offering unnecessarily weak evidence functions. Admissibility is a separate requirement from truthful reporting, and the two can conflict.

To summarize, our requirements are: no benefit (or harm) for participating null agents, maximum benefit for truthful non-null agents, strictly positive net benefit for participating  non-null agents, and admissible evidence functions. Any menu of e-values satisfies the first requirement, but it is clearly a nontrivial task to design menus that satisfy all four. We will do so in this Gaussian setting, as well as in much more general nonparametric settings.

\subsection{Why likelihood-ratio rewards do not solve the problem}

Even a statistically valid contract need not elicit an honest report. With known $\sigma$, the likelihood ratio indexed by a reported mean $m>0$ is
\begin{equation}
e_m(x)=\exp\!\left\{\frac{mx}{\sigma^2}-\frac{m^2}{2\sigma^2}\right\}.
\label{eq:known-lr}
\end{equation}
It is an e-value for $H_0$, but
\begin{equation}
\E_{\mu,\sigma} e_m(X)=\exp\!\left\{\frac{m\mu}{\sigma^2}\right\}.
\label{eq:lr-inflation}
\end{equation}
Every positive-mean agent prefers a larger report. Restricting the report space to a closed bounded interval makes these agents prefer its upper endpoint; it does not separate them by their true means.

Substituting a reported standard deviation introduces a second failure. The candidate
\begin{equation}
e_{m,s}(x)=\exp\!\left\{\frac{mx}{s^2}-\frac{m^2}{2s^2}\right\}
\end{equation}
has, at a true null mean,
\begin{equation}
\E_{0,\sigma}e_{m,s}(X)
=\exp\!\left\{\frac{m^2(\sigma^2-s^2)}{2s^4}\right\}>1
\quad\text{when }s<\sigma.
\label{eq:variance-lr-fail}
\end{equation}
Understating the noise can therefore make an ineffective proposal profitable. Each offered contract must be valid under \emph{all} null variances, including those inconsistent with its report.


\subsection{What this paper establishes}

We construct admissible contracts meeting the validity, truthfulness, and participation requirements in both Gaussian regimes. When the agent knows both parameters, the contract scores three bounded summaries of the observation; the direction of their expectation vector identifies the mean and variance. When the agent knows only its mean, the contract compares a Gaussian kernel centered at the report with the same kernel centered at zero. Its expected reward is uniquely maximized at the true mean regardless of the variance. These results require only one observation. We extend them to sample sizes larger than one and combine elicitation with report-independent evidence rewards, so that honest reporting need not come at the expense of increasing rewards as evidence accumulates.

The non-Gaussian results show both the possibilities and the limits of this approach. Some Gaussian constructions extend to unknown symmetric unimodal distributions. We also consider testing whether a nonnegative distribution has mean below or above one\footnote{This is in fact a hidden encoding of arbitrary testing problems after converting the data to e-values (these are different from the e-value menus provided by the principal: here the data for an arbitrary testing problem can without loss of generality be transformed into e-values)}. Over all such finite-mean distributions, we construct an unbounded contract based on two independent observations that elicits the alternative mean and is individually admissible. Its expected reward depends only on the mean, so the agent needs no information about variance or distributional shape; the variance may be infinite. We prove that one observation cannot achieve the same conjunction of strict mean elicitation and individual admissibility on this class. This is a restriction on \emph{admissible statistical contracts}, not on ordinary mean elicitation, which can use one observation.

A final theme is the distinction between testing and identification. A test that separates nulls from alternatives can yield profitable evidence without identifying the information to be reported. Conversely, a truthful score need not satisfy the null expectation budget or be admissible. Building on composite-testing and constraint-based descriptions of e-values \citep{zhang2024,clerico2025,lrr2026}, we give sufficient constructions and counterexamples that separate these requirements. In particular, boundedness is optional, and a finite description of the null does not by itself guarantee identification or admissible elicitation.

Section~\ref{sec:related} places these results in the broader literature, and Section~\ref{sec:model} states the model precisely. Sections~\ref{sec:full}--\ref{sec:multiple} develop the Gaussian contracts and fixed-sample extensions; Sections~\ref{sec:beyond}--\ref{sec:unbounded} address distributional generality, identification, constraints, and unbounded rewards. 

Throughout, we assume that agent reports precede a verifiable experiment and agents maximize expected realized payouts. Strategic data manipulation is arguably impossible to handle, but handling nonlinear utility, and jointly choosing a report and a stopping rule, are left for future work.

\section{Related work and scope}
\label{sec:related}

\subsection{Strategic testing and principal--agent models}

\citet{tetenov2016} studies regulatory testing when a privately informed agent proposes an innovation. \citet{bates2022} formulate principal--agent hypothesis testing through data-dependent profit licenses and participation incentives, and identify e-value menus as the relevant objects for null deterrence. Their framework is general enough to contain nuisance parameters and multiple reports. Our main difference is the insistence on further properties like strict elicitation and pointwise admissibility, and explicitly understanding how to handle nuisance parameters (that do not matter to the principal and may be unknown to the agent), including some infinite dimensional cases.

\citet{bates2023} study Bayesian inference from participation by a privately informed agent. \citet{shi2024} extend strategic testing to risk-sensitive agents and derive implications for error control and private-information elicitation. These papers make clear that utility, entry decisions, and the content of private information cannot be treated interchangeably. We explicitly impose linear realized payoff and do not claim that the contracts remain truthful under arbitrary concave utility, endogenous costs, or a profit license that cannot be fully realized.

The closest scoring-rule connection is \citet{shi2025}, who develop separating menus for heterogeneous agents with private prior probabilities in a binary-state testing model. Proper scoring rules and report-dependent contracts are central there. Our use of scoring geometry is therefore not a new general connection between elicitation and principal--agent testing. The distinctions are the report target (a sampling-model parameter rather than a prior probability), nuisance-uniform optimality when the agent itself lacks nuisance information, a report-independent participation cost, and admissibility of each e-value. \citet{hossain2025} also analyze strategic hypothesis testing with endogenous participation and reporting, emphasizing how strategic behavior changes threshold design. That design objective is related to, but different from, the strict truthful-report menus constructed here.

\subsection{Proper scores, elicitable functionals, and multiple observations}

The quadratic and spherical geometries used below are classical scoring-rule ingredients. General accounts of proper scoring rules include \citet{gneiting2007} and the recent review of \citet{waghmare2026}. Point-functional elicitation is developed in \citet{gneiting2011}; quantiles, means, and modes do not admit interchangeable scoring rules. Our Gaussian-kernel score elicits the maximizer of a smoothed density.

Finite-dimensional identification and the number of reports required to recover statistical properties are studied systematically by \citet{frongillo2021}. Multiple observations can fundamentally alter elicitation possibilities \citep{casalaina2017}. Our bounded-feature and unbounded finite-mean constructions use that established principle: products of expectations create report geometry unavailable to one-observation linear expectations. The additional constraints here are nonnegative payouts, a common composite-null expectation budget, alternative participation, and admissibility. In particular, the two-observation bounded-mean result below is not a claim that ordinary mean elicitation requires two observations; squared error already elicits a bounded mean with one observation. It is the simultaneous requirement of \emph{e-value admissibility} that produces the one-observation obstruction. The same distinction applies on $[0,\infty)$: one integrable separating e-value can elicit the mean with one observation, but the admissible complete class requires a second independent observation to separate reports.

\subsection{E-values, nuisance parameters, and optimality criteria}

The recent history of e-values is covered in depth in~\cite{ramdas2025hypothesis}.
The most common optimization criterion of \emph{log-optimality} is not the risk-neutral report-choice criterion studied here. 

An e-value can be excellent for testing a specified alternative without making that alternative its uniquely optimal report; most e-values are not ``truthful''. Likewise, our admissibility results do not imply other forms of efficiency or optimality.

\citet{ramdas2022} analyze admissibility in anytime-valid sequential inference and the role of nonnegative martingales. We use the corresponding no-pointwise-improvement notion for a \emph{fixed experimental horizon}. A family containing an admissible e-value at every deterministic horizon need not be an admissible e-process, or even an e-process. We do not delve into e-processes in this paper.

\subsection{Composite separation and hypotheses generated by constraints}

The existence of a common powerful test is itself a nontrivial composite-hypothesis question. \citet{zhang2024} study powerful p-values and e-values for composite nulls and alternatives, including exactness, convex-analytic separation, and the effect of multiple observations. A more recent effort was undertaken by~\cite{larsson2026complete}. Our elementary test-to-score conversion presupposes a \emph{single} test whose power exceeds its null level at every alternative. (It does not assert that every disjoint pair of composite families admits such a test at a given sample size.)
It also concerns positive expected linear payoff, which should be distinguished from positive expected log evidence.

\citet{clerico2025} establish complete-class results for properly constrained hypotheses. \citet{lrr2026} substantially generalize constraint-based descriptions of e-values, including finite measurable constraints and quasi-sure formulations. Their finite-constraint theorem allows unbounded measurable constraints and characterizes feasible evidence through nonnegative combinations of those functions; their admissibility converse has an explicit constraint qualification. We use this representation to expose an elicitation obstruction, with direct arguments for bounded and nonnegative unbounded observations. Our iid product constructions do not claim to characterize all e-values for a product null; this remains distinct from applying a finite marginal-constraint description to all joint distributions. A finite list of null constraints describes a null family, not necessarily the information needed to identify each alternative.

Multiplicative betting and mixtures do not require bounded payouts: nonnegative one-step factors with controlled expectation suffice \citep{vovk2021}. In our nonnegative finite-mean setting, the bets $1+\lambda(X-1)$ are unbounded above but bounded away from zero when $0<\lambda<1$; the first moment controls their expected logarithms. Betting constructions for bounded means are studied by \citet{waudby2024}; joint mean and variance testing by e-processes is developed by \citet{fan2025}. These works concern evidence and inference, not necessarily elicitation of an agent's private sampling parameters. At the opposite extreme, unrestricted mean testing encounters the classical contamination obstruction of \citet{bahadur1956}. We give a finite-sample e-value proof showing why no positive-mean agent can be guaranteed positive expected profit over that unrestricted class.

The results below synthesize these ingredients under a particular set of economic and statistical constraints.

\section{Model, incentive requirements, and admissibility}
\label{sec:model}

Let $\{P_\theta:\theta\in\Theta\}$ be a sampling family, with null types $\Theta_0$ and alternative types $\Theta_1$. For a fixed $n$, data have law $P_\theta^{\otimes n}$. A report $r$ in a report space $\mathcal R$ is chosen before these data are observed. Reports do not change their distribution. The principal offers a measurable nonnegative function $E_{r,n}$ and pays $C_n E_{r,n}$, where the strictly positive, known cost $C_n$ is independent of $r$ and $\theta$.

\subsection{Valid and truthful menus}
We impose no upper bound on $E_{r,n}$. Null validity already implies integrability under each null. To make linear expected-payoff comparisons meaningful, we require
\begin{equation}
\E_\theta E_{r,n}<\infty
\quad\text{for every alternative type $\theta$ and every offered report $r$}.
\label{eq:integrability}
\end{equation}
One can allow extended-valued evidence statistically, but strict elicitation with several infinite expected rewards would be ill-defined. All constructed contracts are finite at every observation and satisfy~\eqref{eq:integrability} on their stated type spaces. Realized upper bounds, uniform expected bounds, and finite expectations under individual types are different assumptions; we identify which is used whenever they matter. E-values in the standard and constraint-based formulations are not required to be bounded above \citep{vovk2021,lrr2026}.

\begin{definition}[Valid and truthful menus]
A menu is \emph{null-valid} if
\begin{equation}
\sup_{\theta\in\Theta_0}\E_\theta E_{r,n}\leq1
\quad\text{for every }r\in\mathcal R.
\label{eq:valid-menu}
\end{equation}
It strictly elicits a target $T(\theta)$ on $\Theta_1$ if
\begin{equation}
\E_\theta E_{T(\theta),n}>\E_\theta E_{r,n}
\quad\text{whenever }r\ne T(\theta).
\label{eq:strict-truth}
\end{equation}
It has \emph{strict alternative participation} if
\begin{equation}
\E_\theta E_{T(\theta),n}>1\quad\text{for every }\theta\in\Theta_1.
\label{eq:participation}
\end{equation}
\end{definition}

Null validity ensures that no null report has positive expected net payoff. Exact boundary-null contracts may leave the agent indifferent between entry and opting out. Where actual nonparticipation is needed, we use opt-out tie-breaking; strict deterrence is stated separately when available. We do not claim strict profitable truth-telling for null types, whose intended action is to decline.

For Gaussian data, the fully informed report is $(\mu,\sigma)$, with report space $(0,\infty)^2$. Reporting standard deviation or variance is equivalent through a known one-to-one transformation. The nuisance-uninformed report space is $(0,\infty)$, with target $\mu$.

\begin{definition}[Nuisance-uniform truthfulness]
A mean-reporting menu is nuisance-uniformly truthful if, for every $\mu>0$,
\begin{equation}
\argmax_{m>0}\E_{\mu,\sigma} E_{m,n}=\{\mu\}
\quad\text{for every }\sigma>0.
\label{eq:nuisance-truth}
\end{equation}
This is a state-by-state expected-payoff statement, not pointwise dominance in the realized data.
\end{definition}

For a proper prior $\Pi$ on nuisance values, state-by-state comparisons integrate to strict comparisons provided the expected payouts of the truthful report and each competing report are integrable under $\Pi$:
\begin{equation}
\int \E_{\mu,\sigma}E_{r,n}\,\Pi(d\sigma)<\infty.
\label{eq:prior-integrability}
\end{equation}
A uniform realized bound is sufficient, but not necessary. A uniform bound on conditional expected payouts for each fixed report also suffices. In particular, when the expected payout depends only on the known mean, every proper prior on distributions with that mean satisfies~\eqref{eq:prior-integrability}, even for unbounded contracts. No variance prior needs to be elicited to choose the optimal report. Gaussian pointwise participation need not give positive \emph{worst-case} surplus over unrestricted variances; in contrast, the finite-mean construction in Section~\ref{sec:unbounded} has the same strictly positive surplus at every shape with a fixed alternative mean.

\subsection{The admissibility notion used here}

Let $\PP_n=\{P_\theta^{\otimes n}:\theta\in\Theta_0\}$. A measurable set is $\PP_n$-polar if every law in $\PP_n$ assigns it probability zero. Equality or dominance outside a polar set is called quasi-sure equality or dominance, abbreviated q.s.

\begin{definition}[Admissible e-value]
A null-valid e-value $E$ is admissible if every null-valid $G$ satisfying $G\geq E$ q.s. also satisfies $G=E$ q.s.
\end{definition}

For the Gaussian family, polar sets are precisely the Lebesgue-null sets. In the nonparametric examples on $[0,1]$ and $[0,\infty)$, every sample point can receive mass under some null product law; admissibility is then pointwise. Upper bounds on the evidence play no role in this definition. This is a complete-class notion for individual evidence functions. It is not a claim that an entire menu is optimal among incentive-compatible mechanisms.

\begin{lemma}[Exactness and coverage]
\label{lem:exact}
Suppose $E$ is valid and there is a subfamily $\PP_n^*\subseteq\PP_n$ such that $\E_P E=1$ for every $P\in\PP_n^*$. If every nonpolar measurable set has positive probability under some $P\in\PP_n^*$, then $E$ is admissible. In particular, one exact null distribution equivalent to a common dominating reference measure suffices.
\end{lemma}
\begin{proof}
If $G\geq E$ and $P(G>E)>0$ for an exact null $P$, then $\E_PG>\E_PE=1$, contradicting validity. Coverage ensures that every nonpolar strict improvement is detected by such a $P$.
\end{proof}

The exactness proof uses nonnegativity and the null expectation budget, not an upper bound on either contract. Exactness at an attained null is sufficient but not necessary. The following elementary observation will be useful when variance is unrestricted.

\begin{lemma}[Diffuse-null exhaustion]
\label{lem:diffuse}
Let the null family be dominated by a measure $\nu$, and suppose it contains laws $P_j$ with densities $p_j$ and constants $a_j>0$ satisfying
\begin{equation}
a_jp_j(x)\longrightarrow w(x)>0\quad\nu\text{-a.e.}
\label{eq:density-exhaust}
\end{equation}
Assume the polar sets coincide with the $\nu$-null sets. A valid $E$ is admissible if
\begin{equation}
a_j\bigl(1-\E_{P_j}E\bigr)\longrightarrow0.
\label{eq:budget-exhaust}
\end{equation}
\end{lemma}
\begin{proof}
For a valid dominator $G$, put $D=G-E\geq0$, choosing nonnegative versions off a null set. Validity gives
\[
0\leq a_j\E_{P_j}D\leq a_j(1-\E_{P_j}E)\longrightarrow0.
\]
Fatou's lemma yields $\int Dw\,d\nu\leq0$. Since $w>0$ a.e., $D=0$ a.e., proving admissibility.
\end{proof}

\begin{corollary}[Zero-integral perturbations under diffuse Gaussian nulls]
\label{cor:zero-integral}
Suppose a dominated null family on $\R^d$ contains $N_d(0,\sigma^2 I_d)$ for all $\sigma>0$ and has Lebesgue-null polar sets. If $E=1+f$ is valid, $f\in L^1(\R^d)$, and $\int f(x)\,dx=0$, then $E$ is admissible.
\end{corollary}
\begin{proof}
Take $a_\sigma=\sigma^d$. The scaled null density tends to $(2\pi)^{-d/2}$ and is bounded by that constant. Dominated convergence gives
\[
\sigma^d\bigl(1-\E_{0,\sigma}E\bigr)
=-(2\pi)^{-d/2}\int f(x)e^{-\norm{x}^2/(2\sigma^2)}\,dx\longrightarrow0.
\]
Apply Lemma~\ref{lem:diffuse} along any sequence $\sigma\to\infty$.
\end{proof}

\section{A Gaussian agent who knows both parameters}
\label{sec:full}

We first construct an admissible one-observation menu that strictly elicits both mean and variance. The key is to identify the type through a \emph{direction} of bounded moments, allowing a spherical rather than a penalized quadratic score.

Fix $a>0$ and $h>0$, with the units of a variance, and let $a_j=a+(j-1)h$, $j=1,2,3$. Define
\begin{equation}
H_j(x)=\frac{x}{\sqrt{a_j}}\exp\!\left\{\frac{1-x^2/a_j}{2}\right\}.
\label{eq:odd-features}
\end{equation}
These functions are odd and lie in $[-1,1]$: the maximum of $t\exp\{(1-t^2)/2\}$ on $t\geq0$ is one at $t=1$.

A Gaussian integration gives, with $v=\sigma^2$,
\begin{equation}
d_j(\mu,\sigma):=\E_{\mu,\sigma}H_j(X)
=\sqrt{\mathrm e}\,a_j\mu(a_j+v)^{-3/2}
  \exp\!\left\{-\frac{\mu^2}{2(a_j+v)}\right\}.
\label{eq:odd-moments}
\end{equation}
For example, completing the square first yields
\[
\E_{\mu,\sigma}e^{-X^2/(2a_j)}
=\sqrt{\frac{a_j}{a_j+v}}\,e^{-\mu^2/(2(a_j+v))},
\]
and the corresponding tilted first moment is $\mu a_j/(a_j+v)$ times this expression. Each $d_j$ has the sign of $\mu$. For $\mu>0$, define the positive unit vector
\begin{equation}
u(\mu,\sigma)=\frac{d(\mu,\sigma)}{\norm{d(\mu,\sigma)}_2}.
\label{eq:direction}
\end{equation}

\begin{lemma}[Three moments identify the positive Gaussian type projectively]
\label{lem:projective}
The map $(\mu,\sigma)\mapsto u(\mu,\sigma)$ is injective on $(0,\infty)^2$.
\end{lemma}
\begin{proof}
Write $A=a+\sigma^2>a$ and $z=\mu^2>0$. The direction determines the ratios
\begin{align}
L_{12}&=\log\frac{d_1/a_1}{d_2/a_2}
 =\frac32\log\frac{A+h}{A}-\frac{zh}{2A(A+h)},\label{eq:L12}\\
L_{23}&=\log\frac{d_2/a_2}{d_3/a_3}
 =\frac32\log\frac{A+2h}{A+h}-\frac{zh}{2(A+h)(A+2h)}.
\label{eq:L23}
\end{align}
Fix $L=L_{12}$. Equation~\eqref{eq:L12} forces
\begin{equation}
z(A)=\frac{2A(A+h)}{h}
\left\{\frac32\log\frac{A+h}{A}-L\right\}.
\label{eq:zA}
\end{equation}
The feasible set $\{A>a:z(A)>0\}$ is an interval, since the expression in braces is strictly decreasing. On this interval the second ratio becomes
\[
F_L(A)=\frac32\log\frac{A+2h}{A+h}
-\frac{A}{A+2h}\left\{\frac32\log\frac{A+h}{A}-L\right\}.
\]
Differentiation and substitution from~\eqref{eq:zA} give
\begin{equation}
F_L'(A)=-\frac{z(A)h^2}{A(A+h)(A+2h)^2}<0.
\label{eq:ratio-derivative}
\end{equation}
Thus $L_{23}$ identifies $A$, after which $L_{12}$ identifies $z$. Positivity selects $\mu=\sqrt z$, and $\sigma=\sqrt{A-a}$.
\end{proof}

For a report $r=(m,s)\in(0,\infty)^2$, choose a fixed $0<\eta<1$ and offer
\begin{equation}
\boxed{\quad
E_r^{\mathrm{sph}}(x)
=1+\frac{\eta}{\sqrt3}\ip{u(m,s)}{H(x)}.
\quad}
\label{eq:spherical-menu}
\end{equation}

\begin{theorem}[Fully informed Gaussian agent]
\label{thm:spherical}
Every contract in~\eqref{eq:spherical-menu} lies in $[1-\eta,1+\eta]$, is valid for all $\mu\leq0$, $\sigma>0$, and is admissible. The menu strictly elicits $(\mu,\sigma)$ on $\mu>0$ and gives every such type strictly positive expected net payoff.
\end{theorem}
\begin{proof}
Cauchy--Schwarz gives $|\ip{u}{H}|\leq\sqrt3$, proving the range. Under the null, all coordinates of $d(\mu,\sigma)$ are nonpositive, while those of $u(m,s)$ are positive. Therefore
\[
\E_{\mu,\sigma}E_r^{\mathrm{sph}}
=1+\frac{\eta}{\sqrt3}\ip{u(r)}{d(\mu,\sigma)}\leq1.
\]
At every centered null this expectation equals one. Such a Gaussian law has an everywhere positive density, so Lemma~\ref{lem:exact} proves admissibility.

For a positive true type $\theta=(\mu,\sigma)$,
\begin{equation}
\E_\theta E_r^{\mathrm{sph}}
=1+\frac{\eta}{\sqrt3}\norm{d(\theta)}_2\ip{u(r)}{u(\theta)}.
\label{eq:spherical-expectation}
\end{equation}
The inner product of unit vectors is at most one, with equality only for equal vectors. Lemma~\ref{lem:projective} makes equality equivalent to $r=\theta$. The truthful surplus before multiplication by $C_1$ is $\eta\norm{d(\theta)}_2/\sqrt3>0$.
\end{proof}

The expected loss from an incorrect report is explicit:
\begin{equation}
\E_\theta(E_\theta^{\mathrm{sph}}-E_r^{\mathrm{sph}})
=\frac{\eta}{\sqrt3}\norm{d(\theta)}_2
 \bigl(1-\ip{u(r)}{u(\theta)}\bigr)>0.
\label{eq:spherical-regret}
\end{equation}
There is no claim that three features are minimal among all possible constructions. They are sufficient for the two-dimensional projective identification needed by this spherical score. A raw moment map being injective would not, by itself, suffice: proportional expectation vectors induce the same normalized vector.

A menu based only on a Student statistic at a fixed sample size cannot strictly distinguish $(\mu,\sigma)$ from $(c\mu,c\sigma)$, $c>0$, because the statistic has the same distribution for both types. The scales $a_j$ in~\eqref{eq:odd-features} deliberately retain absolute-scale information. This is an elicitation requirement, not an argument against invariant tests for their usual testing objectives.

\section{A Gaussian agent who does not know its variance}
\label{sec:unknown}

Fix a bandwidth $a>0$ and $0<\eta<1$. With $K_a(x)=\exp\{-x^2/(2a)\}$, define the menu
\begin{equation}
\boxed{\quad
E_m^{\mathrm{ker}}(x)
=1+\eta\{K_a(x-m)-K_a(x)\},\qquad m>0.
\quad}
\label{eq:kernel-menu}
\end{equation}
The second kernel is a report-independent baseline. It does not change which report maximizes expected reward, but it is essential to the null expectation budget.

\begin{theorem}[Nuisance-uninformed Gaussian agent]
\label{thm:kernel}
Every contract in~\eqref{eq:kernel-menu} lies in $[1-\eta,1+\eta]$, is admissible, and has expectation strictly below one at every finite null type. For every $\mu>0$ and every $\sigma>0$, its expected value is uniquely maximized at $m=\mu$, with truthful expectation strictly greater than one. These strict reporting and participation conclusions also hold under any proper prior on $\sigma\in(0,\infty)$ conditional on a known positive $\mu$.
\end{theorem}
\begin{proof}
The range follows from $0<K_a\leq1$. Put $b=a+\sigma^2$ and $c_\sigma=\sqrt{a/b}$. Gaussian integration yields
\begin{equation}
\E_{\mu,\sigma}E_m^{\mathrm{ker}}
=1+\eta c_\sigma
\left\{e^{-(\mu-m)^2/(2b)}-e^{-\mu^2/(2b)}\right\}.
\label{eq:kernel-expectation}
\end{equation}
If $\mu\leq0$ and $m>0$, then $(\mu-m)^2>\mu^2$, proving strict null validity. If $\mu>0$, the first exponential has its unique maximum at $m=\mu$, independently of $\sigma$. The truthful surplus is
\begin{equation}
\E_{\mu,\sigma}E_\mu^{\mathrm{ker}}-1
=\eta c_\sigma\left(1-e^{-\mu^2/(2b)}\right)>0.
\label{eq:kernel-surplus}
\end{equation}
The report regret is
\begin{equation}
\E_{\mu,\sigma}(E_\mu^{\mathrm{ker}}-E_m^{\mathrm{ker}})
=\eta c_\sigma\left(1-e^{-(m-\mu)^2/(2b)}\right)>0
\quad(m\ne\mu).
\label{eq:kernel-regret}
\end{equation}
Boundedness permits integration of these strict positive differences under any proper prior on finite positive $\sigma$.

For admissibility, $f_m(x)=\eta\{K_a(x-m)-K_a(x)\}$ is integrable and has zero Lebesgue integral, because translation preserves the kernel's integral. Corollary~\ref{cor:zero-integral} applies. More explicitly, the centered-null deficit is
\begin{equation}
1-\E_{0,\sigma}E_m^{\mathrm{ker}}
=\eta\sqrt{\frac a{a+\sigma^2}}
 \left(1-e^{-m^2/[2(a+\sigma^2)]}\right)=O(\sigma^{-3}),
\label{eq:kernel-deficit}
\end{equation}
whereas a nonzero nonnegative local improvement would cost order $\sigma^{-1}$ under sufficiently diffuse centered nulls. Lemma~\ref{lem:diffuse} formalizes this contradiction.
\end{proof}

\begin{remark}[Strict null slack need not imply inadmissibility]
Every finite null variance gives strict slack in~\eqref{eq:kernel-expectation}, yet no fixed positive pointwise improvement is possible. The relevant issue is not whether a particular null attains the budget, but whether the entire null family leaves enough budget for a common nonnegative improvement. Here the asymptotically vanishing budget is exhausted by diffuse nulls.
\end{remark}

\begin{remark}[The principal's agnosticism and the agent's uncertainty]
The principal does not use a variance estimate or a reported variance in this contract. The agent also does not need either, in order to know that $m=\mu$ is uniquely optimal. This is stronger than optimizing expected utility for one specified variance prior. 
\end{remark}

\section{Several observations and rewards for accumulating evidence}
\label{sec:multiple}

Throughout this section $n$ is fixed before reporting and sampling, unless an e-process statement is explicitly made. All admissibility claims refer to the full experiment on $\R^n$.

\subsection{Products preserve fixed-horizon report ordering}

\begin{proposition}[Independent products]
\label{prop:products}
Let $e_r(X)>0$ be an iid one-observation menu with null validity, strict truthfulness, and strict alternative participation. Then
\begin{equation}
A_{r,n}=\prod_{i=1}^n e_r(X_i)
\label{eq:product}
\end{equation}
has the same properties at every deterministic $n$. If $e_r$ is exact at a null law $P_*$ whose $n$-fold product assigns positive probability to every nonpolar set, then $A_{r,n}$ is admissible in the full product experiment. No realized upper bound is needed, but the finite-expectation assumption~\eqref{eq:integrability} is maintained.
\end{proposition}
\begin{proof}
Independence gives $\E_\theta A_{r,n}=(\E_\theta e_r)^n$. Since all one-observation expectations are positive, taking the $n$th power preserves their strict ordering. Exactness and coverage under the product null yield admissibility by Lemma~\ref{lem:exact}.
\end{proof}

Consequently, products of the spherical contracts solve the fully informed problem for every $n$. For fixed reports they also form nonnegative null supermartingales; the same null validity survives predictable, data-dependent report choices. This last observation concerns statistical validity only. The truthful-report theorem concerns a report fixed before the experiment, not a strategic sequence of reports or a jointly optimized stopping rule.

Products of the one-observation kernel contracts retain validity and fixed-horizon truthfulness. We do not infer their full-experiment admissibility merely from one-observation admissibility. Shared nuisance parameters make such an inheritance claim nontrivial. Instead, the next construction gives an admissibility proof directly on $\R^n$.

\subsection{A full-vector kernel for the nuisance-uninformed agent}

Let $\one_n$ be the all-ones vector and define
\begin{equation}
K_{a,n}(x)=\exp\{-\norm{x}_2^2/(2a)\},\qquad x\in\R^n.
\end{equation}
Offer
\begin{equation}
\boxed{\quad
A_{m,n}^{\mathrm{ker}}(x)
=1+\eta\{K_{a,n}(x-m\one_n)-K_{a,n}(x)\},\qquad m>0.
\quad}
\label{eq:vector-kernel}
\end{equation}

\begin{theorem}[Admissible mean elicitation for every fixed sample size]
\label{thm:vector-kernel}
For every fixed $n\geq1$, the menu~\eqref{eq:vector-kernel} is bounded in $[1-\eta,1+\eta]$, admissible, strictly nuisance-uniformly truthful, strictly profitable at every positive mean, and strictly unprofitable at every finite null type.
\end{theorem}
\begin{proof}
Using independence, with $b=a+\sigma^2$,
\begin{equation}
\E_{\mu,\sigma}A_{m,n}^{\mathrm{ker}}
=1+\eta\left(\frac a b\right)^{n/2}
 \left\{e^{-n(\mu-m)^2/(2b)}-e^{-n\mu^2/(2b)}\right\}.
\label{eq:vector-kernel-expectation}
\end{equation}
The same distance comparison as in Theorem~\ref{thm:kernel} proves validity, strict truthfulness, and participation. The perturbation in~\eqref{eq:vector-kernel} is an integrable translation difference on $\R^n$ and has integral zero. Corollary~\ref{cor:zero-integral}, with $d=n$, proves admissibility. Equivalently, its centered-null deficit is $O(\sigma^{-(n+2)})$, which exhausts the scaled null budget after multiplication by $\sigma^n$.
\end{proof}

The vector-kernel score is a fixed-horizon family, not automatically an e-process. Its bounded elicitation surplus can become small as $n$ increases. A report-independent evidence component resolves the latter issue without changing the report ranking.

\subsection{A nuisance-free evidence component}

Set $S_i=\sgn(X_i)$ and define
\begin{equation}
B_n=\int_0^1\prod_{i=1}^n(1+\lambda S_i)\,d\lambda,
\qquad B_0=1.
\label{eq:sign-mixture}
\end{equation}
Gaussian observations have zero probability of a tie at zero. This is a standard mixture-of-bets construction, specialized to a report-independent sign experiment \citep{vovk2021,grunwald2024}.

\begin{proposition}[Validity and growth of the sign mixture]
\label{prop:sign}
The process $(B_n)$ is a nonnegative supermartingale under every Gaussian null. It is bounded by $2^n$ at each fixed $n$ and is exact under every centered Gaussian null. Under every fixed positive-mean Gaussian alternative,
\begin{equation}
\E_{\mu,\sigma} B_n
=\int_0^1(1+\lambda d)^n\,d\lambda>1,
\qquad d=2\Phi(\mu/\sigma)-1>0,
\label{eq:sign-exp}
\end{equation}
and, almost surely,
\begin{equation}
\lim_{n\to\infty}\frac{\log B_n}{n}
=\KL\!\left(\Ber(p)\,\middle\|\,\Ber(1/2)\right)>0,
\qquad p=\Phi(\mu/\sigma).
\label{eq:sign-growth}
\end{equation}
\end{proposition}
\begin{proof}
Under the null, $\E S_i\leq0$, so each product at a fixed $\lambda\in[0,1]$ is a nonnegative supermartingale. Tonelli's theorem gives the same property for the mixture. If $\mu=0$, each factor has conditional expectation one, giving exactness. Independence also proves~\eqref{eq:sign-exp}.

For growth, let $N_n=\sum_{i=1}^n\ind\{X_i>0\}$ and put $q=(1+\lambda)/2$. Then
\[
B_n=2\int_{1/2}^1(2q)^{N_n}\{2(1-q)\}^{n-N_n}\,dq.
\]
Almost surely $N_n/n\to p\in(1/2,1)$. On every closed interval inside $(1/2,1)$, the normalized log integrand converges uniformly to
\[
\ell(q)=p\log(2q)+(1-p)\log\{2(1-q)\}.
\]
This strictly concave function is maximized at $q=p$, with maximum equal to the divergence in~\eqref{eq:sign-growth}. Integrating over a small interval around $p$ gives the matching lower exponential bound. For the upper bound, bound the integral by its largest integrand; eventually its maximum on $[1/2,1]$ is at $q=N_n/n$. Taking limits proves~\eqref{eq:sign-growth}.
\end{proof}

For example, when all observations are positive,
\begin{equation}
B_n=\int_0^1(1+\lambda)^n\,d\lambda
=\frac{2^{n+1}-1}{n+1}.
\end{equation}
The score does not use magnitudes, so it is not proposed as a Gaussian-efficiency optimum. It gives a particularly transparent strong-evidence component with no report or nuisance dependence.

\subsection{Combining elicitation with evidence rewards}

Choose a fixed $0<\gamma<1$ and define
\begin{equation}
\boxed{\quad
\widetilde E_{r,n}=(1-\gamma)A_{r,n}+\gamma B_n.
\quad}
\label{eq:combined}
\end{equation}
Use a product of spherical scores for $A_{r,n}$ in the fully informed regime, and the vector-kernel score in the nuisance-uninformed regime.

\begin{theorem}[Truthful, admissible contracts with growing evidence rewards]
\label{thm:combined}
For either information regime, every fixed-$n$ contract in~\eqref{eq:combined} is admissible and null-valid. The menu strictly elicits the agent's specified information and gives every positive-mean type strictly positive expected profit. Under any fixed positive-mean Gaussian law, $\widetilde E_{r,n}\to\infty$ almost surely as $n\to\infty$, for any sequence of admissible reports $r$ at the respective horizons.
\end{theorem}
\begin{proof}
Convex combinations preserve validity. Since $B_n$ is report-independent, expected report differences equal $(1-\gamma)$ times the corresponding differences for $A_{r,n}$. Both truthful components have expectation above one under each alternative, proving participation.

For the spherical product, both components are exact at every centered null, so Lemma~\ref{lem:exact} gives admissibility. For the vector kernel, $\E_{0,\sigma}B_n=1$ and hence
\[
\sigma^n\bigl(1-\E_{0,\sigma}\widetilde E_{m,n}\bigr)
=(1-\gamma)\sigma^n\bigl(1-\E_{0,\sigma}A_{m,n}^{\mathrm{ker}}\bigr)\longrightarrow0.
\]
Lemma~\ref{lem:diffuse} applies directly. Finally $\widetilde E_{r,n}\geq\gamma B_n$, so almost-sure growth follows from Proposition~\ref{prop:sign}, independently of reports.
\end{proof}

For the mean-only construction, every finite null type strictly prefers opting out; for the spherical construction, centered nulls are indifferent. For these Gaussian constructions, payouts are bounded at each deterministic $n$, so arbitrary proper nuisance priors cause no integrability problem. This is a property of the constructions, not a restriction on the framework. With payment $C_n\widetilde E_{r,n}$, the evidence-to-cost ratio is unbounded along alternative sample paths. Absolute payments also diverge if $\inf_n C_n>0$. The claims compare different predetermined sample sizes; they do not promise arbitrarily large payouts from one fixed bounded experiment.

Neither mixing nor admissibility settles incentives under agent-chosen optional stopping. The spherical product mixture is a null e-process; the vector-kernel mixture need not be. Even when a process is null-valid at stopping times, the report maximizing a stopped expected payoff can differ from the one maximizing every deterministic-horizon expectation.

\section{Beyond Gaussian distributions}
\label{sec:beyond}

Two different routes go beyond Gaussian variance nuisance parameters. The first retains a symmetric unimodal location structure and uses bounded kernel scores, as developed here. The second imposes only a known lower support bound and a finite first moment, and uses unbounded evidence; its full construction appears in Section~\ref{sec:unbounded}.

\subsection{Unknown symmetric unimodal shape}

The mean-only kernel construction uses a location property that is not specific to Gaussian variance calculations. Let
\begin{equation}
X=\mu+Z,
\label{eq:location-general}
\end{equation}
where $Z$ has an unknown symmetric unimodal density $f$: $f(z)=f(-z)$ and $f$ is nonincreasing on $[0,\infty)$. We impose no common variance, tail bound, or parametric shape. When the first absolute moment is finite, $\mu=\E X$; otherwise $\mu$ is a location or median parameter rather than a finite mean.

\begin{lemma}[Strict peak after Gaussian smoothing]
\label{lem:smoothing}
For any symmetric unimodal probability density $f$ and $a>0$, the function
\begin{equation}
g(t)=\int_\R K_a(z-t)f(z)\,dz
\end{equation}
is positive, even, and strictly decreasing on $(0,\infty)$.
\end{lemma}
\begin{proof}
Evenness follows by symmetry. By the layer-cake representation, $f$ is an integral of indicators of centered intervals, up to null sets. For any finite interval radius $r>0$, its convolution contribution is
\[
J_r(t)=\int_{-r}^{r}K_a(z-t)\,dz.
\]
For $t>0$,
\[
J_r'(t)=K_a(r+t)-K_a(r-t)<0,
\]
since $|r+t|>|r-t|$ and the Gaussian kernel is strictly decreasing in absolute argument. Integrability of $f$ makes the positive superlevel intervals finite for almost every positive level, and a positive set of levels has positive radius. Integrating the strict comparison $J_r(t_2)<J_r(t_1)$ for $0\leq t_1<t_2$ proves strict decrease. Tonelli's theorem justifies the nonnegative integral representation.
\end{proof}

\begin{theorem}[Shape-uninformed location elicitation]
\label{thm:unimodal}
For iid data from~\eqref{eq:location-general}, the vector-kernel menu~\eqref{eq:vector-kernel} is strictly truthful for the positive location $\mu$, uniformly over the unknown symmetric unimodal density. It is null-valid for $\mu\leq0$, strictly profitable at every positive location, and admissible relative to this density class. With a finite first moment, these statements apply to the mean.
\end{theorem}
\begin{proof}
Independence and symmetry give
\begin{equation}
\E_{\mu,f}A_{m,n}^{\mathrm{ker}}
=1+\eta\{g(\mu-m)^n-g(\mu)^n\}.
\label{eq:unimodal-expectation}
\end{equation}
Lemma~\ref{lem:smoothing} makes the first term uniquely maximal at $m=\mu$. For a null $\mu\leq0$ and $m>0$, $|\mu-m|>|\mu|$, giving strict null validity. Positive-location participation follows from $g(0)>g(\mu)$. The null class contains all centered Gaussian scales and is dominated by Lebesgue measure; its polar sets are Lebesgue-null. The zero-integral perturbation argument therefore proves admissibility also for this larger null class.
\end{proof}

The sign mixture can again be added. A symmetric unimodal density gives $\Pr(X>0)>1/2$ whenever $\mu>0$, so its evidence grows at a positive rate. At $\mu=0$ it is exact for every continuous symmetric law. Consequently the validity, reporting, admissibility, and growth proofs of Theorem~\ref{thm:combined} extend to this shape-uninformed setting.

Symmetry alone is not enough for the kernel score: a symmetric multimodal density can have smoothed modes away from its center. The theorem's unimodality condition is therefore substantive, not merely convenient notation.

\subsection{Arbitrary continuous medians: truthfulness without admissibility}

An alternative mechanism handles any continuous distribution with a unique median. Fix $w(t)>0$ on $[0,\infty)$ with $\int_0^\infty w(t)\,dt=1$ and let $W(m)=\int_0^m w(t)\,dt$. Define
\begin{equation}
E_m^{\mathrm{med}}(x)
=1+\eta\int_0^m w(t)\{2\ind\{x>t\}-1\}\,dt,
\quad m>0,\quad0<\eta<1.
\label{eq:median-score}
\end{equation}
For a continuous distribution function $F$,
\begin{equation}
\frac{d}{dm}\E_FE_m^{\mathrm{med}}
=\eta w(m)\{1-2F(m)\}.
\end{equation}
If $F$ has a unique median $\mu>0$, this derivative is positive below $\mu$ and negative above it. For nonpositive medians the expected integral is nonpositive, while at a positive truthful report it is strictly positive. Thus the mechanism is valid, strictly truthful, and strictly profitable for the positive median. This is consistent with classical quantile scoring theory \citep{gneiting2011}.

However, it is not generally admissible. The valid sign contract
\begin{equation}
E_m^+(x)=1+\eta W(m)\sgn(x)
\label{eq:median-dominator}
\end{equation}
pointwise dominates~\eqref{eq:median-score}, with strict improvement on $0<x<m$. Under a centered Gaussian, this interval has positive probability, so inadmissibility holds even for the Gaussian submodel. The improved score has expectation $1+\eta W(m)\E\sgn(X)$: every positive-median agent now prefers larger $m$. Improving individual e-values can destroy the incentive properties of the menu. This also shows why a construction that works for all unknown symmetric shapes need not inherit the kernel construction's admissibility.

\subsection{Unrestricted finite-mean distributions}

\begin{proposition}[No universally profitable alternative with unrestricted tails]
\label{prop:unrestricted}
Fix $n<\infty$. Let the null include every distribution on $\R$ with finite mean at most zero. For any valid nonnegative $E$ on $\R^n$ and any iid alternative law $P$ with finite positive mean, $\E_{P^{\otimes n}}E\leq1$. Consequently no valid menu can offer strict participation to such an alternative.
\end{proposition}
\begin{proof}
Let $\mu=\E_PX>0$. For $\eps\in(0,1)$ define
\[
Q_\eps=(1-\eps)P+\eps\delta_{z_\eps},\qquad
z_\eps=-\frac{(1-\eps)\mu}{\eps}.
\]
This is a finite-mean null with mean zero. Nonnegativity gives
\[
1\geq\E_{Q_\eps^{\otimes n}}E
\geq(1-\eps)^n\E_{P^{\otimes n}}E.
\]
Let $\eps\downarrow0$.
\end{proof}

This is the e-value form of the contamination phenomenon associated with \citet{bahadur1956}; related one-observation consequences also follow from constraint representations \citep{lrr2026}. Gaussianity is not necessary, but some restriction preventing the arbitrarily negative contamination used here is necessary for the stated universal mean-participation goal. The proof never uses an upper bound on $E$, so allowing unbounded rewards does not evade it. A known lower support bound supplies such a restriction without imposing an upper tail bound, as Section~\ref{sec:unbounded} demonstrates.

\section{From composite tests to elicitation}
\label{sec:tests}

We now consider arbitrary families $\PP$ and $\QQ$ of laws on a \emph{fixed} experiment $\X$. These may themselves be product-experiment laws. The useful interpretation of ``there is a test'' is the existence of one test with power above its null level at every alternative. The bounded-test equivalence below is one sufficient route, not a characterization of all integrable separating e-values.

\begin{proposition}[Bounded tests and positive-payoff e-values]
\label{prop:test-equiv}
The following are equivalent:
\begin{enumerate}[label=(\roman*),nosep]
\item There exist $0<\alpha<1$ and $\varphi:\X\to[0,1]$ such that $\sup_{P\in\PP}\E_P\varphi\leq\alpha$ and $\E_Q\varphi>\alpha$ for every $Q\in\QQ$.
\item There exist $M>1$ and an e-value $e:\X\to[0,M]$ for $\PP$ such that $\E_Qe>1$ for every $Q\in\QQ$.
\end{enumerate}
\end{proposition}
\begin{proof}
For (i)$\Rightarrow$(ii), use $e=\varphi/\alpha$ and $M=1/\alpha$. For (ii)$\Rightarrow$(i), use $\varphi=e/M$ and $\alpha=1/M$. A level-zero test with positive alternative power can instead be converted by $e=1+c\varphi$, $c>0$.
\end{proof}

This is an elementary rescaling, not a new existence theorem. The existence of a common separating test is studied more deeply by \citet{zhang2024}. Pointwise power above size does not imply a uniform positive power gap. Pairwise distinguishability, with a different test for each pair of distributions, does not supply the common test required in Proposition~\ref{prop:test-equiv}.

Given the bounded e-value, put
\begin{equation}
H(x)=\frac{e(x)-1}{\max\{1,M-1\}}\in[-1,1].
\label{eq:separator}
\end{equation}
Then $\E_PH\leq0$ under the null and $d_Q:=\E_QH>0$ under every alternative. The score
\begin{equation}
S(q,x)=\frac{2qH(x)-q^2}{3},\qquad0\leq q\leq1,
\label{eq:bounded-quadratic}
\end{equation}
lies in $[-1,1]$ and satisfies
\begin{equation}
\E_QS(q,X)=\frac{d_Q^2-(q-d_Q)^2}{3}.
\end{equation}
Thus it strictly elicits $d_Q$, and $1+\tau S(q,x)$ is a nonnegative valid e-value for $0<\tau<1$. At the truthful $q=d_Q$, its expectation exceeds one.

The precise answer to the proposed conversion is therefore \emph{yes for an expectation of the testing statistic}. It is not automatically a score for the desired mean, variance, or distribution. Different alternatives may have the same $d_Q$; if the agent does not know a nuisance parameter, it may not know $d_Q$ even when it knows the target parameter. For a score that is already bounded, the interval $[-1,1]$ is a convenient normalization. The existence of a bounded separator, however, is a genuine restriction, as the next two subsections show. Neither normalization nor an upper bound establishes identification or removes nuisance dependence.

\subsection{Direct elicitation from an integrable, possibly unbounded e-value}

\begin{proposition}[Spherical elicitation of an e-value expectation]
\label{prop:unbounded-conversion}
Let $e\geq0$ be an e-value for $\PP$, and suppose
\[
q(Q):=\E_Qe\in(1,\infty)\qquad(Q\in\QQ).
\]
For reports $r>1$, define
\begin{equation}
\boxed{\quad
F_r(x)=\frac{1+r e(x)}{\sqrt{2(1+r^2)}}.
\quad}
\label{eq:unbounded-conversion}
\end{equation}
Then $(F_r)$ is a nonnegative, integrable, null-valid menu that strictly elicits $q(Q)$ and gives each alternative strictly positive expected surplus. An upper bound on $e$ is unnecessary. Each contract in this particular conversion is inadmissible.
\end{proposition}
\begin{proof}
For a null $P$,
\[
\E_PF_r\leq\frac{1+r}{\sqrt{2(1+r^2)}}<1\qquad(r>1),
\]
and nonnegativity is immediate. For $q=\E_Qe>1$,
\begin{equation}
\E_QF_r=\frac{1+rq}{\sqrt{2(1+r^2)}}
\leq\sqrt{\frac{1+q^2}{2}},
\label{eq:unbounded-conversion-exp}
\end{equation}
with equality exactly at $r=q$, by Cauchy--Schwarz and the leading constant coordinate. The maximum exceeds one, and all expectations are finite. Finally,
\[
\delta_r=1-\frac{1+r}{\sqrt{2(1+r^2)}}>0
\]
is a uniform null-budget slack. The nonnegative contract $F_r+\delta_r$ is valid and pointwise strictly larger, proving inadmissibility.
\end{proof}

This is the same classical spherical geometry used elsewhere in the paper, now applied to $(1,e)$. It avoids a potentially invalid attempt to shift an arbitrary score with no finite lower bound. Its target is still $q(Q)$, not automatically the mean or the full type. If $q(Q)$ depends on an unknown nuisance, the uninformed agent need not know which report is optimal. Removing boundedness therefore removes neither the identification question nor the admissibility question.

\subsection{Bounded separation is not without loss of generality}

\begin{proposition}[An unbounded separator with no bounded counterpart]
\label{prop:no-bounded-separator}
On the one-observation space $[0,\infty)$, let
\begin{equation}
\PP_+=\{P:\E_PX\leq1\},\qquad
\QQ_+=\{Q:1<\E_QX<\infty\}.
\label{eq:nonnegative-families}
\end{equation}
The function $e(x)=x$ is integrable and separates every alternative in expected payoff. There is no bounded e-value $f$ for $\PP_+$ with $\E_Qf>1$ for every $Q\in\QQ_+$.
\end{proposition}
\begin{proof}
The claims for $e(x)=x$ follow from the definitions. Let $f$ be any bounded valid e-value. The null $\delta_0$ gives $f(0)\leq1$. If $f(0)<1$, then for $M>2$ the alternative
\[
Q_M=(1-2/M)\delta_0+(2/M)\delta_M
\]
has mean two and satisfies $\E_{Q_M}f\to f(0)<1$. It therefore violates the desired strict separation for large $M$.

If $f(0)=1$, apply null validity at the mean-one law
\[
P_x=(1-1/x)\delta_0+(1/x)\delta_x,\qquad x>1.
\]
It implies $f(x)\leq1$. For $x\leq1$, validity at $\delta_x$ gives the same inequality. Thus $f\leq1$ everywhere and separates no alternative.
\end{proof}

This does not contradict Proposition~\ref{prop:test-equiv}, whose second statement explicitly requires bounded evidence. Thresholding any e-value yields valid level tests, but positive \emph{expected linear payoff} under an alternative does not guarantee that one common threshold test has power above its level under every alternative. The composite existence results of \citet{zhang2024} distinguish boundedness and their particular notions of power; our example concerns the linear-payoff criterion used here.

Truncation illustrates the quantifier issue. For each fixed alternative with $\E_Qe>1$, monotone convergence gives $\E_Q(e\wedge M)>1$ for all sufficiently large $M$. A single truncation suffices for finitely many alternatives. For an infinite class, its required level may depend on the alternative; Proposition~\ref{prop:no-bounded-separator} proves that no common level, or any other bounded separator, need exist.

There is a related limitation at every finite sample size, even though the preceding no-separator claim was only for one observation.
\begin{proposition}[Bounded contracts cannot give shape-uniform positive surplus]
\label{prop:no-uniform-bounded}
Fix $n<\infty$ and an alternative mean $\mu>1$. If $E$ is a bounded e-value for $\{P^{\otimes n}:P\in\PP_+\}$, then
\begin{equation}
\inf_{Q:\,\E_QX=\mu}\E_{Q^{\otimes n}}E\leq1.
\label{eq:no-uniform-bounded}
\end{equation}
\end{proposition}
\begin{proof}
Let $Q_M=(1-\mu/M)\delta_0+(\mu/M)\delta_M$ for $M>\mu$. The probability that its iid sample contains a nonzero observation tends to zero. Boundedness gives $\E_{Q_M^{\otimes n}}E\to E(0,\ldots,0)\leq1$, where the last inequality follows from null validity at $\delta_0$.
\end{proof}

For a mean-reporting menu, this applies to the single truthful contract $E_\mu$. It rules out a strictly positive surplus uniformly over all shapes with a fixed mean, not pointwise strict participation for every shape. We do not claim that bounded pointwise separators are impossible for every $n>1$. The unbounded construction in Section~\ref{sec:unbounded} achieves the stronger, shape-uniform surplus in~\eqref{eq:no-uniform-bounded} with the inequality reversed strictly.

\subsection{Finite separating moments and their limitations}

A vector version is immediate. Suppose $H_j\in[-1,1]$, $j=1,\ldots,k$, have nonpositive null expectations and nonnegative alternative expectations. For positive weights $w_j$ summing to one, let $d_j(Q)=\E_QH_j$ and offer
\begin{equation}
E_r(x)=1+\lambda\sum_{j=1}^kw_j
 \{2d_j(r)H_j(x)-d_j(r)^2\},\qquad0<\lambda<1/3.
\label{eq:general-quadratic}
\end{equation}
This is bounded, nonnegative, and null-valid, and
\begin{equation}
\E_QE_r
=1+\lambda\norm{d(Q)}_w^2-\lambda\norm{d(r)-d(Q)}_w^2.
\label{eq:general-quadratic-regret}
\end{equation}
An injective, nonzero $d$ gives strict full-type elicitation and participation. For nuisance-uninformed agents, the optimal report must be computable from the information they possess; nuisance-dependent moments do not generally meet that requirement.

Finitely many such moments are sufficient for this recipe, not necessary for every truthful menu. Countably many bounded features with summable positive weights can support the same argument, provided they identify the target and have the stipulated signs. Conversely, simple injectivity of a finite vector is insufficient for the \emph{spherical} recipe, which requires identification after normalization. Admissibility adds a further constraint. With the stated scaling, every nonzero-report contract in~\eqref{eq:general-quadratic} is dominated by the valid nonnegative contract $1+2\lambda\sum_jw_jd_j(r)H_j(x)$, obtained by removing its strictly positive quadratic penalty. Thus this direct quadratic conversion is inadmissible. 

There is also no reason that every identifying feature must itself be a separating test. An independent observation allows us to use one separator to satisfy the null budget and other features to elicit the report.

\subsection{A general independent-observation construction}

\begin{theorem}[A separating statistic times an identifying spherical score]
\label{thm:gate}
Suppose $H:\X\to[-1,1]$ satisfies
\[
a(P):=\E_PH\leq0\quad(P\in\PP),\qquad
 a(Q)>0\quad(Q\in\QQ).
\]
Let $G:\X\to[0,1]^k$ be bounded features and $t(P)=\E_PG$. A report $r$ specifies $t(r)\in[0,1]^k$. For independent $X,Y\sim P$ and $0<\eta<1$, define
\begin{equation}
\boxed{\quad
E_r(X,Y)=1+\frac{\eta}{\sqrt{k+1}}H(X)
 \frac{1+\ip{t(r)}{G(Y)}}{\sqrt{1+\norm{t(r)}_2^2}}.
\quad}
\label{eq:gate}
\end{equation}
This is a valid e-value in $[1-\eta,1+\eta]$. Under every $Q\in\QQ$, its expected value is uniquely maximized in the encoded report at $t(r)=t(Q)$ and has truthful expectation above one. If $t$ identifies the desired report, the menu strictly elicits that report. If $t(P)$ depends only on a known target parameter while $a(P)>0$ may depend on an unknown nuisance, truthfulness is nuisance-uniform.

If there is a family of null laws with $a(P)=0$ whose product laws cover every nonpolar set, every contract in~\eqref{eq:gate} is admissible.
\end{theorem}
\begin{proof}
By Cauchy--Schwarz,
\[
0<\frac{1+\ip{t(r)}{G(Y)}}{\sqrt{1+\norm{t(r)}_2^2}}
\leq\sqrt{1+\norm{G(Y)}_2^2}\leq\sqrt{k+1},
\]
proving the range. Independence yields
\begin{equation}
\E_PE_r
=1+\frac{\eta a(P)}{\sqrt{k+1}}
 \frac{1+\ip{t(r)}{t(P)}}{\sqrt{1+\norm{t(r)}_2^2}}.
\label{eq:gate-expectation}
\end{equation}
The fraction is positive, so a nonpositive $a(P)$ proves null validity. Under an alternative, Cauchy--Schwarz bounds that fraction by $\sqrt{1+\norm{t(P)}_2^2}$. Equality holds exactly when $(1,t(r))$ is proportional to $(1,t(P))$. The leading coordinate one forces the proportionality factor to equal one, hence $t(r)=t(P)$. The truthful surplus is
\begin{equation}
\frac{\eta a(Q)}{\sqrt{k+1}}\sqrt{1+\norm{t(Q)}_2^2}>0.
\end{equation}
The multiplier $a(P)$ does not affect the maximizer when it is positive, proving the nuisance-uniform statement. Exactness when $a(P)=0$ and Lemma~\ref{lem:exact} prove admissibility under the stated coverage condition.
\end{proof}

The role of independence is explicit: it converts the expectation of the product in~\eqref{eq:gate} into the product of expectations in~\eqref{eq:gate-expectation}. The two observations can be independent data blocks, each as large as the experiment needed for the common test. This changes the sampling budget and does not claim a free one-observation conversion. Averaging the contract with its version having $X$ and $Y$ exchanged gives a symmetric contract with the same expected values and, under the same exactness condition, the same admissibility.

The theorem refines the finite-feature intuition. With an independent observation, it is sufficient to have \emph{one} common bounded separator plus bounded identifying moments, not a finite collection all of whose coordinates individually separate null and alternative. But neither an identifying moment representation nor the exact-null coverage needed for this admissibility proof follows from the mere existence of a test. The boundedness assumptions are local to this sufficient construction: replacing $G$ by an arbitrary unbounded feature could make $H(X)$ times the score arbitrarily negative. Section~\ref{sec:unbounded} instead exploits nonnegative observations and a bounded negative part to maintain nonnegative payouts without an upper bound.

\section{Finite constraints and the bounded-mean example}
\label{sec:constraints}

\subsection{What the constraint representation implies}

Let $g=(g_1,\ldots,g_k)$ be real measurable constraints and consider the full distribution class
\begin{equation}
\PP=\{P:\E_P|g_j|<\infty,\ \E_Pg_j\leq0,\ j=1,\ldots,k\}.
\label{eq:finite-null}
\end{equation}
Theorem~3.1 of \citet{lrr2026} says that every e-value is q.s. dominated by a nonnegative function of the form
\begin{equation}
1+\ip{\pi}{g(x)},\qquad \pi\in[0,\infty)^k.
\label{eq:constraint-complete}
\end{equation}
Their Corollary~3.3 consequently puts every admissible e-value in this class. 
Earlier complete-class results under regular finite constraints are due to \citet{clerico2025}. These representation results are not restricted to bounded constraints or bounded e-values; on $[0,\infty)$ the single unbounded constraint $g(x)=x-1$ is already covered.

The implication for reporting is immediate. For menus within~\eqref{eq:constraint-complete}, alternative preferences have the form
\begin{equation}
\E_Q E_r=1+\ip{\pi(r)}{t(Q)},\qquad t(Q)=\E_Qg.
\label{eq:linear-preferences}
\end{equation}

\begin{proposition}[Positive-ray obstruction]
\label{prop:ray}
If two alternative types satisfy $t(Q_2)=c\,t(Q_1)$ for some $c>0$, then they have identical rankings over every menu of the form~\eqref{eq:constraint-complete}. Such a menu cannot uniquely elicit distinct reports for those two types.
\end{proposition}
\begin{proof}
For any two reports, their expected payoff difference under $Q_2$ is $c$ times that under $Q_1$.
\end{proof}

Thus even an injective raw constraint expectation does not guarantee strict elicitation: direction, not magnitude, determines rankings in this affine class. Finite-constraint theory characterizes feasible e-values; it does not automatically make the constraint expectations a suitable strictly elicitable type representation.

\subsection{One observation on \texorpdfstring{$[0,1]$}{[0,1]}: an impossibility theorem}

Fix $c\in(0,1)$ and let $\PP_c$ contain all laws on $[0,1]$ with mean at most $c$. Let $\QQ_c$ contain all such laws with mean greater than $c$. The complete class below is a special case of the cited constraint theorem; we include an elementary proof to make the elicitation consequence transparent. In this particular one-observation model, validity itself forces every e-value to be bounded. For $x>c$, the mean-$c$ null $(1-c/x)\delta_0+(c/x)\delta_x$ and nonnegativity give $e(x)\leq x/c\leq1/c$; for $x\leq c$, $e(x)\leq1$ follows from $\delta_x$. Thus dropping a payout cap cannot remove this model's obstruction.

\begin{proposition}[Complete class for the bounded-mean null]
\label{prop:bounded-complete}
Every one-observation e-value for $\PP_c$ is pointwise dominated by
\begin{equation}
e_\lambda(x)=1+\lambda(x-c),\qquad 0\leq\lambda\leq1/c.
\label{eq:affine-bounded}
\end{equation}
Every function in~\eqref{eq:affine-bounded} is admissible. Hence these are exactly the admissible one-observation e-values.
\end{proposition}
\begin{proof}
Let $e$ be valid. For $x\leq c$, the null point mass at $x$ gives $e(x)\leq1$. For $x<c<y$, the two-point law with mean $c$ gives
\[
(y-c)e(x)+(c-x)e(y)\leq y-x.
\]
Equivalently,
\begin{equation}
\frac{e(y)-1}{y-c}\leq\frac{1-e(x)}{c-x}.
\label{eq:slope-bound}
\end{equation}
Set
\[
L=\max\!\left\{0,\sup_{y>c}\frac{e(y)-1}{y-c}\right\},\qquad
U=\inf_{x<c}\frac{1-e(x)}{c-x}.
\]
Then $0\leq L\leq U\leq(1-e(0))/c\leq1/c$. The upper bound also ensures finiteness. Any $\lambda\in[L,U]$ satisfies $e(x)\leq1+\lambda(x-c)$ for all $x$, including $x=c$.

Conversely,~\eqref{eq:affine-bounded} is nonnegative and has expectation at most one under the null. It has expectation one under every law with mean $c$. Each point of $[0,1]$ can have positive probability under such a law, by adding an appropriate endpoint mass if necessary. Lemma~\ref{lem:exact} therefore proves pointwise admissibility. A valid $e$ is admissible only if it equals its affine dominator.
\end{proof}

\begin{theorem}[Admissible one-observation contracts cannot elicit a positive bounded mean]
\label{thm:bounded-impossible}
With a report-independent cost and linear payoff, no menu of individually admissible one-observation e-values for $\PP_c$ can strictly elicit the mean at two different values in $(c,1]$.
\end{theorem}
\begin{proof}
By Proposition~\ref{prop:bounded-complete}, every report offers a coefficient $\lambda(r)$ and has expected payoff proportional to
\[
1+\lambda(r)(\mu-c).
\]
Every $\mu>c$ strictly prefers a larger coefficient. All positive-mean types therefore share the same set of optimal reports; if a maximum coefficient is not attained, none has an optimal report. Distinct strictly truthful reports are impossible.
\end{proof}

The null here has only one bounded constraint, and its expectation $\mu-c$ is nonzero and identifies the desired positive mean. Nevertheless admissibility and strict one-observation elicitation are incompatible. This directly rules out the claim that finitely many identifying constraints alone guarantee an admissible truthful menu.

Dropping admissibility removes this particular obstruction: the bounded quadratic construction in Section~\ref{sec:tests}, with $H(x)=(x-c)/\max\{c,1-c\}$, gives a valid truthful one-observation menu. Theorem~\ref{thm:bounded-impossible} is not an impossibility result for mean elicitation or for valid truthful menus without admissibility.

\subsection{Two observations: an explicit admissible solution}

Let $b=\max\{c,1-c\}$, choose $0<\eta<1$, and let the agent report $m\in(c,1]$. For iid $X,Y\in[0,1]$, offer
\begin{equation}
\boxed{\quad
E_m(X,Y)
=1+\frac{\eta}{b\sqrt2}(X-c)\frac{1+mY}{\sqrt{1+m^2}}.
\quad}
\label{eq:bounded-two}
\end{equation}

\begin{theorem}[Two-observation bounded-mean elicitation]
\label{thm:bounded-two}
For the full class of iid distributions on $[0,1]$, every contract in~\eqref{eq:bounded-two} is bounded in $[1-\eta,1+\eta]$, null-valid for $\E X\leq c$, and pointwise admissible. For every mean $\mu>c$, the unique optimal report is $m=\mu$, independently of variance and all other features of the distribution, and truthful expected payoff exceeds one.
\end{theorem}
\begin{proof}
Apply Theorem~\ref{thm:gate} with $H(x)=(x-c)/b$, $G(y)=y$, and $t(m)=m$. In detail,
\begin{equation}
\E_PE_m
=1+\frac{\eta(\mu-c)}{b\sqrt2}
 \frac{1+m\mu}{\sqrt{1+m^2}}.
\label{eq:bounded-two-expectation}
\end{equation}
For $\mu>c$, the derivative of the report-dependent part has the sign of
\[
\frac{\mu-m}{(1+m^2)^{3/2}},
\]
so the unique maximizer is $m=\mu$, including the endpoint case $\mu=1$. The truthful surplus is
\begin{equation}
\frac{\eta(\mu-c)}{b\sqrt2}\sqrt{1+\mu^2}>0.
\end{equation}
The range and null validity follow from the gate theorem.

For admissibility, the contract is exact under every iid law with mean $c$. Fix any point $(x,y)\in[0,1]^2$. A mean-$c$ law can assign positive mass to both $x$ and $y$: allocate sufficiently small positive masses there, then distribute the remaining mass between zero and one to attain mean $c$. The product law assigns positive mass to $(x,y)$. Consequently these exact nulls detect every pointwise strict improvement, and Lemma~\ref{lem:exact} applies.
\end{proof}

The contract may be symmetrized by averaging~\eqref{eq:bounded-two} with its swapped version. This preserves the expected payoff, bounds, validity, and admissibility. Neither version needs a variance report or estimate.

The gain comes from iid replication. The expectation contains a product of moments, $(\mu-c)(1+m\mu)$, unavailable to an affine one-observation expectation. The joint null $\{P^{\otimes2}:\E_PX\leq c\}$ is not the class of all joint distributions satisfying two marginal mean inequalities. Applying a one-observation finite-constraint complete class on $[0,1]^2$ after replacing the iid null by that larger class would be a different problem. This distinction is consistent with the explicit separation between generic constraint representations and product-null characterization in \citet{lrr2026}, and with the role of replication in \citet{casalaina2017}.

\subsection{Mean and variance on bounded support}

The same independent-observation construction can elicit both first and second moments. Let
\[
G(y)=(y,y^2),\qquad
 t(m,v)=(m,m^2+v),
\]
where $m\in(c,1]$ and $0\leq v\leq m(1-m)$. With $H(x)=(x-c)/b$, define
\begin{equation}
E_{m,v}(X,Y)
=1+\frac{\eta}{b\sqrt3}(X-c)
\frac{1+mY+(m^2+v)Y^2}{\sqrt{1+m^2+(m^2+v)^2}}.
\label{eq:bounded-mean-var}
\end{equation}
Theorem~\ref{thm:gate} gives strict elicitation of $(\mu,\operatorname{Var}X)$ for every positive-mean type who knows both, without specifying any remaining distributional shape. Since $t(m,v)$ identifies the pair, distinct parameter reports have strictly different expected payoffs. Exactness at every mean-$c$ null and the same atomic coverage argument prove pointwise admissibility.

Independent products across disjoint pairs preserve truthful report ordering and are exact at every boundary null; they are consequently admissible in the larger iid experiment. Extra unused observations do not change this coverage argument. A report-independent bounded-mean betting mixture based on $H(X_i)$ can be added just as in Section~\ref{sec:multiple}. Under $\E H>0$, a sufficiently small positive bet has positive expected logarithmic growth because $H$ is bounded and the derivative of $\E\log(1+\lambda H)$ at $\lambda=0$ equals $\E H$. A mixture with positive mass near such a bet therefore grows exponentially, and remains exact under boundary means. Thus the bounded-support mechanisms also allow substantial evidence rewards without sacrificing reporting incentives.

\subsection{Why finitely many null constraints do not settle identification}

Two logically separate difficulties remain in the general finite-constraint setting. First, a common separator need not exist for the entire complement of the null in a fixed experiment. On the two-point sample space $\{a,b\}$, let
\[
g(a)=(1,-2),\qquad g(b)=(-2,1).
\]
The point masses $\delta_a$ and $\delta_b$ both violate a constraint and are alternatives, whereas their equal mixture satisfies both constraints. A test with power above $\alpha$ at both point masses would have null expectation above $\alpha$ at their mixture. No such common test exists, despite the finite description and individually identifiable violations.

Second, finitely many bounded moments need not identify an unrestricted alternative distribution. Given any $k$ bounded features on $(c,1]$, choose $k+2$ distinct support points. Their augmented vectors $(1,G(x))\in\R^{k+1}$ are linearly dependent. A sufficiently small signed perturbation of strictly positive weights along this dependence gives two different probability distributions with the same feature expectations. Both have mean greater than $c$ because their supports lie in $(c,1]$. Thus no finite feature vector identifies the whole alternative law on this class. It can still identify a finite target such as a mean or a mean--variance pair, as the constructions above exploit.

The appropriate conclusion is therefore conditional. A common bounded test supplies one possible separating expectation; bounded identifying moments and independent sampling provide one sufficient route to report separation; and exactness or another exhaustion argument can establish admissibility. Integrable unbounded separators and scores provide additional routes, as the next section shows. Finitely many constraints do not automatically provide separation, identification, a suitable sampling structure, and admissibility simultaneously.

\section{Unbounded admissible contracts for nonnegative means}
\label{sec:unbounded}

We now drop the upper support bound, symmetry, unimodality, and any variance assumption. Let $X,Y$ be iid from an arbitrary law $P$ on $[0,\infty)$ with finite mean $\mu$. The null and alternative are the families in~\eqref{eq:nonnegative-families}. The agent knows $\mu$ and reports $m>1$ before sampling; the rest of the distribution is an unknown nuisance. This section gives a complete admissible construction for that information regime. In particular, knowledge or existence of the variance is unnecessary.

The change from $\R$ to a known lower-bounded support is substantive: it prevents the negative contamination used in Proposition~\ref{prop:unrestricted}. We retain arbitrary upper tails. Classical spherical scoring geometry and the multi-observation representation of products of expectations motivate the formulas \citep{gneiting2007,casalaina2017}; the proofs verify the additional null-budget, participation, and admissibility requirements directly.

\subsection{Two observations and a finite first moment}

Define the nonnegative statistics
\begin{equation}
U_1=\frac{X+Y}{2},\qquad U_2=XY.
\label{eq:unbounded-features}
\end{equation}
Independence and Tonelli's theorem give $\E_PU_1=\mu$ and $\E_PU_2=\mu^2<\infty$. The second identity uses only the first moment of each observation, not $\E_PX^2$.

Fix $0<\eta<1/\sqrt2$, let $D_m=\sqrt{1+(m+1)^2}$, and offer
\begin{equation}
\boxed{\quad
E_m^+(X,Y)
=1+\eta\frac{(U_1-1)+(m+1)(U_2-1)}{D_m},
\qquad m>1.
\quad}
\label{eq:unbounded-pair}
\end{equation}

\Needspace{10\baselineskip}
\begin{theorem}[Admissible mean elicitation without a variance assumption]
\label{thm:unbounded-pair}
For iid nonnegative finite-mean observations, every contract~\eqref{eq:unbounded-pair} is nonnegative, integrable, and valid under every null mean $\mu\leq1$. It is pointwise admissible, unbounded above, and exact under every mean-one law. For each $\mu>1$, the unique optimal report is $m=\mu$ under every distribution with that mean. The truthful expected surplus, before multiplying by the participation cost, is
\begin{equation}
\eta(\mu-1)\sqrt{1+(\mu+1)^2}>0.
\label{eq:unbounded-surplus}
\end{equation}
Thus strict truthfulness and strictly positive surplus are uniform over all distributional shapes with the same alternative mean, including infinite-variance shapes.
\end{theorem}
\begin{proof}
Expanding the contract gives
\begin{equation}
E_m^+
=1-\eta\frac{m+2}{D_m}
 +\frac{\eta}{2D_m}(X+Y)
 +\frac{\eta(m+1)}{D_m}XY.
\label{eq:unbounded-coefficients}
\end{equation}
Since $(m+2)/D_m\leq\sqrt2$, all displayed coefficients are nonnegative and $E_m^+\geq1-\eta\sqrt2>0$. Its positive $XY$ coefficient makes it unbounded above. Each term has a finite expectation when $\mu<\infty$, and
\begin{align}
\E_PE_m^+
&=1+\eta\frac{(\mu-1)+(m+1)(\mu^2-1)}{D_m}\nonumber\\
&=1+\eta(\mu-1)
 \frac{1+(m+1)(\mu+1)}{\sqrt{1+(m+1)^2}}.
\label{eq:unbounded-expectation}
\end{align}
For $0\leq\mu\leq1$, the increment is nonpositive; it is zero at every mean-one law and strictly negative when $\mu<1$.

For $\mu>1$, Cauchy--Schwarz gives
\begin{equation}
\frac{1+(m+1)(\mu+1)}{\sqrt{1+(m+1)^2}}
\leq\sqrt{1+(\mu+1)^2},
\label{eq:unbounded-cs}
\end{equation}
with equality exactly at $m=\mu$. This proves strict truthfulness and~\eqref{eq:unbounded-surplus}. Equivalently, the derivative of the fraction in~\eqref{eq:unbounded-expectation} is $(\mu-m)/[1+(m+1)^2]^{3/2}$.

For admissibility, fix any point $(x,y)\in[0,\infty)^2$. Choose $\eps>0$ small enough that $2\eps<1$ and $\eps(x+y)<1$, and set
\[
z=\frac{1-\eps(x+y)}{1-2\eps},\qquad
P_* =\eps\delta_x+\eps\delta_y+(1-2\eps)\delta_z.
\]
This is a nonnegative, finitely supported law with mean one. Its product law gives positive probability to $(x,y)$ and is exact for $E_m^+$. Hence any strict pointwise improvement at that point violates the expectation budget under $P_*^{\otimes2}$. Lemma~\ref{lem:exact} proves pointwise admissibility.
\end{proof}

The expected payout in~\eqref{eq:unbounded-expectation} depends only on $\mu$ and $m$. Thus every proper prior on distributions with the known mean gives exactly the same finite expected payoff, rather than merely preserving its maximizer. Even maximin expected-payoff evaluation over all such shapes gives the positive surplus~\eqref{eq:unbounded-surplus}. This is stronger than the Gaussian kernel's pointwise-in-variance participation claim. It does not provide a uniform gap as $\mu\downarrow1$, uniform testing power, or uniformly rapid evidence growth.
For example, a scaled Pareto law with shape parameter in $(1,2)$ has a finite first moment and an infinite second moment; the elementary tail integrals establish both facts. It is covered by the theorem. 

The expectation increments have the revealing form
\begin{equation}
(\E U_1-1,\E U_2-1)=(\mu-1)(1,\mu+1).
\label{eq:unbounded-projective}
\end{equation}
Their direction identifies $\mu>1$. This is the projective information absent from the one-observation scalar mean constraint. The nonnegative statistics have a bounded negative part after centering, permitting the spherical normalization without bounding the positive rewards. Proposition~\ref{prop:no-uniform-bounded} shows that the shape-uniform positive surplus achieved here cannot be matched by a bounded truthful contract at any fixed sample size.

\subsection{Why two observations are necessary for admissible elicitation}

\begin{proposition}[One-observation complete class on the half-line]
\label{prop:halfline-complete}
Every one-observation e-value for $\PP_+$ is pointwise dominated by
\begin{equation}
e_\lambda(x)=1+\lambda(x-1),\qquad0\leq\lambda\leq1.
\label{eq:halfline-affine}
\end{equation}
These functions are exactly the admissible one-observation e-values. Every nonconstant member is unbounded above.
\end{proposition}
\begin{proof}
Apply the slope argument from Proposition~\ref{prop:bounded-complete} with $c=1$, allowing $y>1$ to range over the full half-line. Two-point mean-one nulls give
\[
\frac{e(y)-1}{y-1}\leq\frac{1-e(x)}{1-x}
\quad(0\leq x<1<y).
\]
The lower and upper slope bounds yield a common $\lambda\in[0,1]$ dominating $e$ on both sides of one; $e(1)\leq1$ handles the remaining point. Each affine function is nonnegative and valid and is exact under every mean-one law. Every point can receive mass under a mean-one law, proving admissibility by Lemma~\ref{lem:exact}. This is also the specialization of the finite-constraint characterization to $g(x)=x-1$ \citep{lrr2026}.
\end{proof}

\begin{corollary}[Sharp sampling threshold]
\label{cor:two-minimal}
Among iid fixed-sample menus with report-independent cost, linear payoff, and individually admissible contracts, two observations are necessary and sufficient to strictly elicit all alternative means $\mu>1$ on the class of nonnegative finite-mean laws while guaranteeing positive alternative surplus.
\end{corollary}
\begin{proof}
With one observation, every admissible report offers a coefficient $\lambda(r)$ and has expected value $1+\lambda(r)(\mu-1)$. All alternatives have identical report rankings, so two distinct means cannot have distinct unique optimal reports. Theorem~\ref{thm:unbounded-pair} supplies the two-observation solution.
\end{proof}

This is not an impossibility for one-observation elicitation without admissibility. Taking $e(x)=x$ in Proposition~\ref{prop:unbounded-conversion} gives a valid, strictly truthful, profitable one-observation menu on the same class, but its contracts have removable null slack. The one-observation obstruction therefore persists after unbounded payouts are allowed and is specifically about their admissible subclass.

\subsection{All fixed sample sizes and strong-evidence rewards}

For iid $X_1,\ldots,X_n$ with $n\geq2$, use the symmetric nonnegative statistics
\begin{equation}
U_{1,n}=\frac1n\sum_{i=1}^nX_i,\qquad
U_{2,n}=\binom n2^{-1}\sum_{i<j}X_iX_j,
\label{eq:unbounded-U}
\end{equation}
and replace $U_1,U_2$ in~\eqref{eq:unbounded-pair} by $U_{1,n},U_{2,n}$ to obtain $E_{m,n}^+$. Their expectations are again $\mu$ and $\mu^2$, so the same nonnegativity, integrability, truthful-report ordering, and surplus proofs apply. Exactness holds under every mean-one product law. Such laws detect every sample point: assign sufficiently small positive masses to all coordinates of that point and use one additional nonnegative atom to attain mean one. Every $E_{m,n}^+$ is therefore pointwise admissible in the full $n$-observation experiment.

Its expected payout does not grow with $n$, since its identifying moment vector does not change. A report-independent mixture supplies evidence accumulation. Let $\lambda_j=2^{-j}$ and $w_j=2^{-j}$, $j\geq1$, and define
\begin{equation}
B_n^+=\sum_{j=1}^\infty w_j
 \prod_{i=1}^n\{1+\lambda_j(X_i-1)\},\qquad B_0^+=1.
\label{eq:unbounded-betting}
\end{equation}
The weights sum to one and $\lambda_j\leq1/2$. Each realized factor is positive and at most $\max\{1,X_i\}$, so the mixture is finite for every finite observation vector.

\begin{proposition}[First-moment evidence growth]
\label{prop:unbounded-growth}
The process $(B_n^+)$ is a nonnegative supermartingale under every iid null in $\PP_+$ and a martingale under every mean-one law. Under every fixed alternative with $1<\mu<\infty$,
\begin{equation}
1<\E_P B_n^+
=\sum_{j=1}^\infty w_j\{1+\lambda_j(\mu-1)\}^n<\infty
\quad(n\geq1),
\label{eq:unbounded-betting-exp}
\end{equation}
and $\liminf_{n\to\infty}n^{-1}\log B_n^+>0$ almost surely. Neither statement requires a finite variance.
\end{proposition}
\begin{proof}
The conditional expectation of each factor is $1+\lambda_j(\mu-1)$, at most one under the null and equal to one at its boundary. Tonelli's theorem gives validity and exactness of the mixture, and independence gives~\eqref{eq:unbounded-betting-exp}. Its upper bound is $\{1+(\mu-1)/2\}^n$ for alternatives.

Write $Z=X-1\geq-1$ and $L(\lambda)=\E\log(1+\lambda Z)$ for $0<\lambda\leq1/2$. The logarithm is integrable: its negative part is at most $\log2$, while its positive part is at most $\lambda Z^+$. Moreover,
\[
\left|\frac{\log(1+\lambda Z)}{\lambda}\right|
\leq Z^+ +2Z^-,
\]
so dominated convergence gives $L(\lambda)/\lambda\to\E Z=\mu-1>0$ as $\lambda\downarrow0$. Some $j$ therefore satisfies $L(\lambda_j)>0$. The strong law and the lower bound by that single mixture component give
\[
\liminf_{n\to\infty}\frac1n\log B_n^+
\geq\lim_{n\to\infty}\left\{\frac{\log w_j}{n}
+\frac1n\sum_{i=1}^n\log(1+\lambda_j(X_i-1))\right\}
=L(\lambda_j)>0.
\]
\end{proof}

\Needspace{10\baselineskip}
\begin{theorem}[Admissible unbounded rewards at every fixed horizon]
\label{thm:unbounded-combined}
For $0<\gamma<1$ and every fixed $n\geq2$, the menu
\begin{equation}
\widetilde E_{m,n}^+=(1-\gamma)E_{m,n}^+ +\gamma B_n^+
\label{eq:unbounded-combined}
\end{equation}
is nonnegative, integrable, null-valid, pointwise admissible, strictly truthful for $\mu>1$, and strictly profitable at every alternative. Its expected payout depends only on the true mean and reported mean. Under each fixed alternative, its evidence grows without bound almost surely as $n$ increases, for any sequence of reports at the predetermined horizons.
\end{theorem}
\begin{proof}
Both components are valid and exact under every mean-one product law. Exact-null coverage proves admissibility, and the report-independent second component preserves the strictly truthful ordering of the first. Their expectations are finite functions of the mean alone and exceed one at truth, proving participation and nuisance-prior integrability. Finally, $\widetilde E_{m,n}^+\geq\gamma B_n^+$ and Proposition~\ref{prop:unbounded-growth} gives the claim about growing horizons.
\end{proof}

The evidence component is an e-process, but the full symmetric elicitation family is not claimed to be an e-process. Fixed-horizon truthfulness and admissibility do not settle joint optimization over reports and stopping times. These unbounded fixed-horizon contracts can already pay arbitrarily large rewards within one experiment; increasing $n$ supplies the additional almost-sure growth property. An externally imposed liability cap would be a new economic constraint and could alter both truthfulness and the shape-uniform surplus.

\subsection{Other thresholds and known lower bounds}

If $X\geq L$ almost surely and the null is $\E X\leq c$ for known $L<c$, transform the data by
\begin{equation}
Z=\frac{X-L}{c-L}\geq0.
\label{eq:lower-bound-transform}
\end{equation}
The null becomes $\E Z\leq1$. Apply the preceding menus to iid transformed observations and encode a reported original mean $m>c$ by $(m-L)/(c-L)$. This one-to-one transformation preserves report incentives, validity, integrability, and admissibility. No upper support or variance bound is introduced. The family of all signed finite-mean distributions on $\R$ has no such common lower bound and remains subject to Proposition~\ref{prop:unrestricted}.

\section{Discussion}
\label{sec:discussion}

The Gaussian constructions show that nuisance validity, truthful elicitation, alternative participation, and admissibility are compatible under linear expected rewards. They do so by two different mechanisms. The fully informed score spends its entire null budget at every centered Gaussian and uses a projectively identifying moment map. The variance-uninformed score is a baseline-subtracted kernel whose preferred location does not depend on variance, with admissibility enforced by diffuse nulls rather than an attained exact null. The latter mechanism extends to unknown symmetric unimodal density shapes.

The bounded-support and nonnegative finite-mean examples show why admissibility should not be appended casually to a truthfulness argument. One-observation admissible evidence has only a single favorable direction, so all positive-mean types prefer the same contract. Independent observations create a strictly separating report geometry while retaining exact null expectations. This is a more informative answer than treating a finite number of constraints as a generic sufficient condition for truthful mechanism design. On the half-line, the one-to-two-observation transition is sharp even when unbounded payouts are allowed. The two-observation menu needs only a finite mean because iid cross-products estimate its square without requiring a second moment.



The strong-evidence mixtures establish growing reward ratios as the experiment expands, and the half-line construction also allows unbounded payouts within a fixed experiment. 
A useful next mathematical objective is optimization \emph{within} the admissible truthful class: maximize a specified evidence or welfare criterion subject to the exact report-ordering restrictions. The present paper establishes feasibility and sharp structural contrasts, rather than such an optimization theorem. 

\paragraph{AI disclosure.} The author acknowledges the extensive use of GPT to develop several threads of initial ideas, and also for organizing the writing and finding connections to related work.

\begingroup
\small
\setlength{\bibsep}{4pt}

\endgroup

\end{document}